\documentclass[pdflatex,sn-mathphys-num]{sn-jnl}

\usepackage{graphicx}%
\graphicspath{{Figures/}}
\usepackage{braket}
\usepackage{amsmath}
\usepackage{caption}
\usepackage{ragged2e}

\theoremstyle{thmstyleone}%

\theoremstyle{thmstyletwo}%

\theoremstyle{thmstylethree}%

\begin{document}

\title{Near-Resonance-Induced Caustics and Scaling Laws in a Quantum Kicked Rotor}

\author[1]{\fnm{Yi} \sur{Cao}}\email{caoyi24@gscaep.ac.cn}

\author[1]{\fnm{Shaowen} \sur{Lan}}\email{lanshaowen22@gscaep.ac.cn}

\author[1]{\fnm{Bin} \sur{Sun}}\email{sunbin20@gscaep.ac.cn}

\author*[1]{\fnm{Jie} \sur{Liu}}\email{jliu@gscaep.ac.cn}

\affil[1]{\orgdiv{Graduate School}, \orgname{China Academy of Engineering Physics}, \city{Beijing}, \postcode{100193}, \country{China}}

\abstract{In this study, we investigate the dynamics of the quantum kicked rotor in the near-resonant regime and observe distinct caustic structures, such as recurring cusps, cusp oscillations, and reticular cusp patterns in high-order resonant cases. By deriving a path integral expression for the wave function's time evolution, we analytically determine both the positions of the caustic singularities and their recurrence periods. We further derive and validate a power-law scaling with an Arnold index of $1/4$, which establishes a quantitative relationship between the amplification of the wave amplitude, the kicking strength, and the resonant detuning parameter. We also explore the classical-quantum correspondence of these caustic singularities, demonstrating that chaos disrupts phase matching and ultimately erodes the caustic structure. Finally, we address the feasibility of experimental implementations of our findings and their broader ramifications for related research fields.}

\keywords{Quantum kicked rotor,~quantum caustics,~catastrophe theory}

\maketitle

\section{Introduction}
The propagation of light through media gives rise to a diverse range of intriguing optical phenomena resulting from ray-focusing effects. Among these, caustics such as glories and rainbows stand out due to their striking visual manifestations and underlying physical mechanisms~\cite{Berry1980, Arnold1983, Stavroudis2012, Berry2015}. These caustic phenomena are by no means confined to optical systems. Across other physical systems, generalized caustics arise as multiple particle trajectories converge in configuration space, giving rise to singularities that induce dramatic amplification of observable effects~\cite{Kravtsov1983, Garg2022, Xia2018, Liao2022, Dong2024}. Notably, in wave catastrophe, caustics manifest as highly localized wave density peaks at phase singularities~\cite{Saha2025, Mumford2019, Kirkby2019, Kirkby2022}, where classical ray-focusing is modulated by quantum interference~\cite{Berry1999, Dell2012, Goldberg2019}. Mathematically, the enhancement of observables follows a universal scaling law derived from catastrophe theory, where the scaling exponent is determined by the Arnold index based on the classification of the catastrophe singularity~\cite{Thom1975, Arnold1976, Arnold1992, Kojakhmetov2024}.

On the other hand, the kicked rotor, as a paradigmatic model of Floquet systems, serves as a fundamental platform for exploring quantum transport, energy diffusion, and nonlinear dynamics in classical-quantum correspondence~\cite{Felix1990, Santhanam2022}. Classically, the system exhibits chaotic diffusion in momentum space, while its quantum counterpart (i.e., the quantum kicked rotor, QKR) induces richer dynamics, including Anderson localization~\cite{Fishman1982, Raizen1999, Billy2008} and quantum resonances~\cite{Casati1979, Izrailev1980, Moore1995}. The formal simplicity and experimental feasibility of the QKR model render it highly valuable for investigating periodically-driven quantum dynamics~\cite{Creffield2006, Wang2009, Chen2014, Bakman2019, Sun2025}. Recently, the QKR model has been extended to nonlinear \cite{Zhang2004, Liu2006, zhaowenlei2014, zhaowenlei2016, Shi2025} and non-Hermitian \cite{zhaowenlei2020, zhaowenlei2022, zhaowenlei2023, zhaowenlei2024} regimes.

The QKR dynamics show notable parallels with optical systems, in which the kicking potential can be effectively modeled as periodic optical gratings (or thin lenses)~\cite{Nye1999, Hecht2016}. This analogy is further supported by previous studies, which have demonstrated that a single delta-pulse-excited quantum rotor generates characteristic caustic structures analogous to the focusing effect of optical thin lenses~\cite{Averbukh2001, Leibscher2002, Mumford2017}. Building on the optical-QKR similarity, the well-established optical caustic formation mechanism~\cite{Berry1976, Berry1980} suggests that analogous structural features may emerge in the QKR phase space, via a combination of catastrophe and chaos theory. This theoretical correspondence allows the QKR to bridge optical caustics and nonlinear dynamical systems. Investigating caustics in the QKR may therefore provide valuable insights into the universal mechanisms underlying the transition from regularity to chaos.

In this paper, we examine the dynamics of a quantum rotor under a sequence of delta kicks and identify caustic structures that exhibit remarkable spatiotemporal periodicity in the near-resonant regime. We develop a comprehensive path integral framework, based on the stationary phase approximation, to provide rigorous analytical validation of these numerical findings. We further discuss the implications of these results and their relevance to experimental realization. This paper is organized as follows. Sec.~\ref{sec-2} details the dynamical evolution of the QKR via numerical simulations. Sec.~\ref{sec-3} offers a theoretical interpretation of the observed recurring cusp caustics, and Sec.~\ref{sec-4} summarizes our conclusions.

\section{Recurring Cusp Caustics in QKR}\label{sec-2}
\begin{figure*}[t]
\centering
\includegraphics[width=1\linewidth]{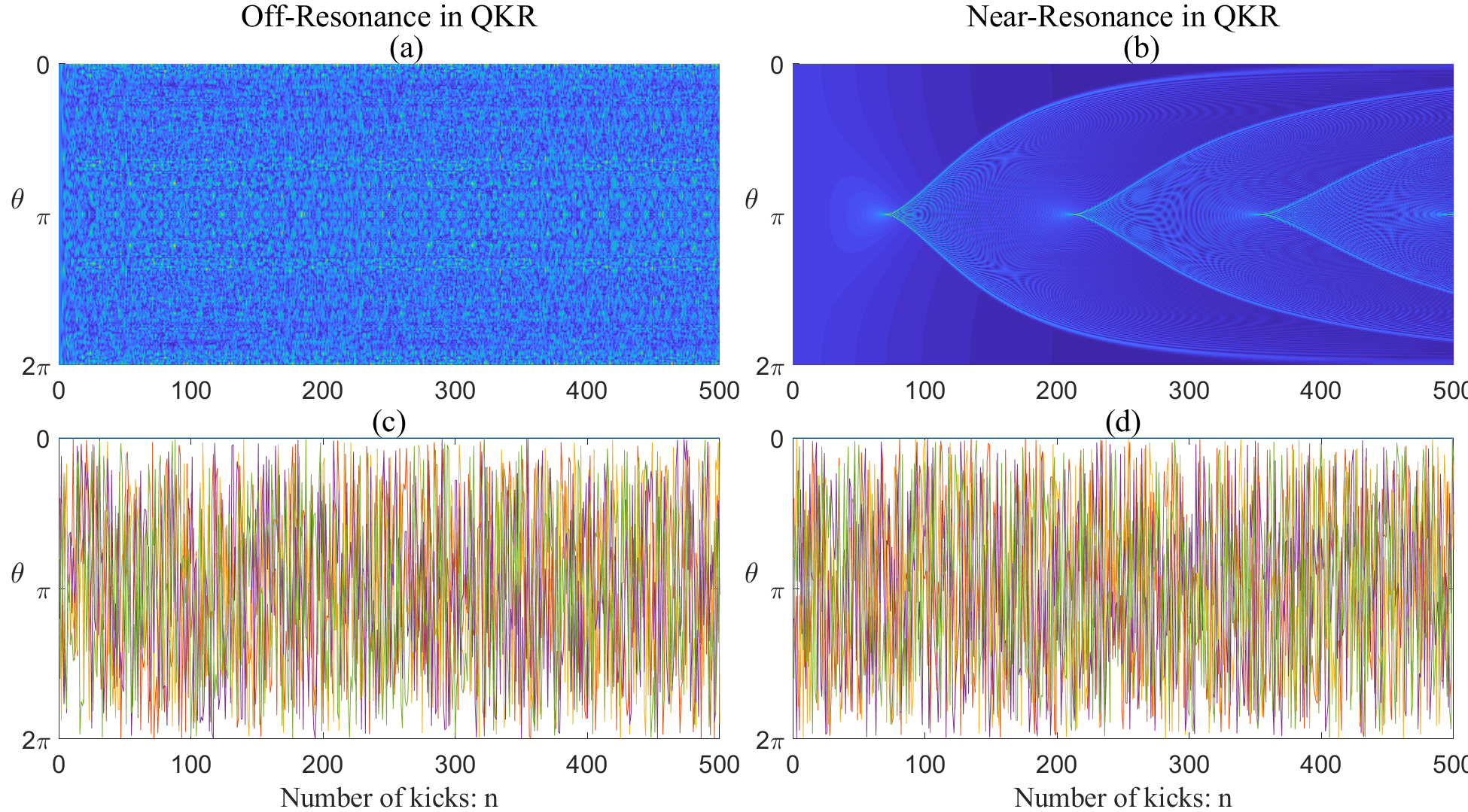}
\caption{Dynamical evolution of the QKR (top) and its classical counterpart (bottom). The quantum system is initialized in the zero-momentum plane-wave state $\psi(\theta_0,t_0)=1/\sqrt{2\pi}$, while classical trajectories are uniformly sampled with initial momentum $p_{0}=0$. Left and right panels correspond to $\Delta=\sqrt{2}$ and $\Delta=0.0001$, respectively. Other parameters are fixed at $K=5$ and $T=4\pi+\Delta$. The wave amplitude $|\psi(\theta_{n},t_{n})|$ is displayed in panels (a) and (b).}
\label{Fig-Quantum Caustics and the Standard Mapping}
\end{figure*}
The dimensionless Hamiltonian of the QKR is $\hat{H}=\hat{p}^2/2+K\cos\hat{\theta}\sum_{n=1}^{\infty}\delta(t-nT)$, where $\hat{\theta}$ denotes the angle operator with eigenvalues $\theta$ lying within $[0,2\pi)$, and $\hat{p}$ is the angular momentum operator. Owing to the periodic boundary conditions imposed on $\theta$, the eigenvalues of $\hat{p}$ are integer-quantized (with $\hbar=1$). Here, $K$ and $T$ denote the kicking strength and period, respectively. The system exhibits primary quantum resonances when the kicking period satisfies $T=4N\pi$, where $N$ is a positive integer~\cite{Santhanam2022}. In this work, we focus on the near-resonant regime and therefore introduce a detuning parameter $\Delta$ from the primary resonant condition by setting the kicking period to $T=4\pi+\Delta$, where $\Delta/4\pi\ll 1$.

In the classical limit, the system's dynamics can be analyzed using the Poincar\'e surface of section, reducing the continuous-time evolution to a discrete-time mapping. Specifically, the evolution follows the mapping: $\theta_{j+1}=\theta_{j}+p_{j}T~(\text{mod}~2\pi),~p_{j+1}=p_{j}+K\sin\theta_{j+1}$, where $(p_{j},\theta_{j})$ denote the angular momentum and angle immediately after the $j$-th kick. Under the momentum rescaling $\tilde{p}_{j}=p_{j}T$, this mapping becomes equivalent to the standard mapping~\cite{Chirikov1979}, characterized by an effective kicking strength $K_{\text{eff}}=KT$. The system undergoes a transition from regular to globally chaotic dynamics when $K_{\text{eff}}$ exceeds the critical value $K_{\text{c}}\approx 0.9716$~\cite{Chirikov1979, Greene1979}. Remarkably, in the quantum resonant regime, phase coherence strongly suppresses chaotic diffusion even when $K_{\text{eff}}>K_{\text{c}}$~\cite{Lepers2008}.

To investigate detuning effects, we perform numerical simulations of the QKR and its classical counterpart (discrete-time mapping) for varying detuning at fixed kicking strength $K=5$. The QKR is initialized in a uniform plane-wave state, $\psi(\theta_0,t_0)=(1/\sqrt{2\pi})e^{ip_{0}\theta_{0}}$, with the zero-momentum case ($p_0=0$) shown in Fig.~\ref{Fig-Quantum Caustics and the Standard Mapping}. In contrast, the classical dynamics are sampled from five trajectories with initial angles uniformly distributed over $[0,2\pi)$ and all at the same initial momentum $p_0$.

For large detuning $\Delta=\sqrt{2}$ as shown in Figs.~\ref{Fig-Quantum Caustics and the Standard Mapping}(a) and (c), the classical dynamics displays global chaos since the effective kicking strength $K_{\text{eff}}$ in the standard mapping substantially exceeds the critical value $K_{\text{c}}$. In the quantum case, the off-resonant condition induces a dynamical competition between dispersion-induced wave packet spreading and kick-driven phase modulation, which disrupts regular quantum evolution, resulting in quantum chaotic dynamics.

In the near-resonant regime $\Delta=0.0001$, as illustrated in Figs.~\ref{Fig-Quantum Caustics and the Standard Mapping}(b) and (d), a pronounced divergence emerges between quantum and classical dynamics. Although $K_{\text{eff}} \gg K_{\text{c}}$ still holds, the classical system remains chaotic, whereas the quantum evolution manifests a regular pattern of recurring cusp caustics. This behavior differs markedly from the caustics observed under single-pulse excitation~\cite{Mumford2017}. Under single-pulse conditions, time-translation symmetry during free evolution decouples the pulse timing from the dynamical evolution, restricting the system to forming only a single cusp caustic. The subsequent free evolution then displays distinct Talbot interference features. Furthermore, even under periodic kicks, no caustics are observed at quantum resonance. Instead, the system exhibits optical Talbot-like interference when the initial wave packet is Gaussian~\cite{Lepers2008}. Interestingly, recurring cusp caustic structures have also been observed in the two-mode Bose-Hubbard model, which is a time-independent interacting many-body system~\cite{Kirkby2022}. In contrast, the QKR system involves a periodically driven external potential that is typically expected to be non-integrable and chaotic.

\begin{figure}[t]
\centering
\includegraphics[width=1\linewidth]{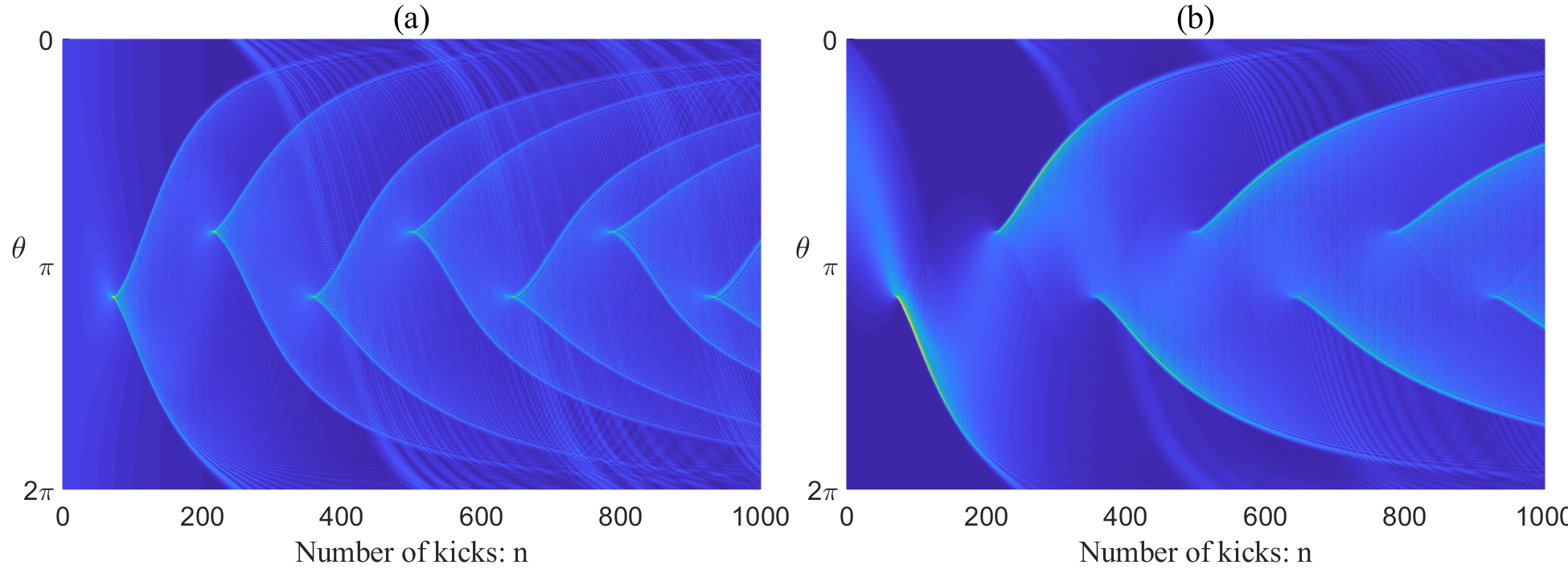}
\caption{Evolution of the wave amplitude $|\psi(\theta_{n},t_{n})|$ for the QKR with $K=5$ and $\Delta=0.0001$. The system is initialized in: (a) a plane-wave state; (b) a Gaussian wave packet in angular space with a standard deviation of 1 and centered at $\pi/2$. The initial momentum of all states is $p_0=100$.}
\label{Fig-Quantum Caustics with Non-zero Initial Momentum}
\end{figure}

\begin{figure}[t]
\centering
\includegraphics[width=1\linewidth]{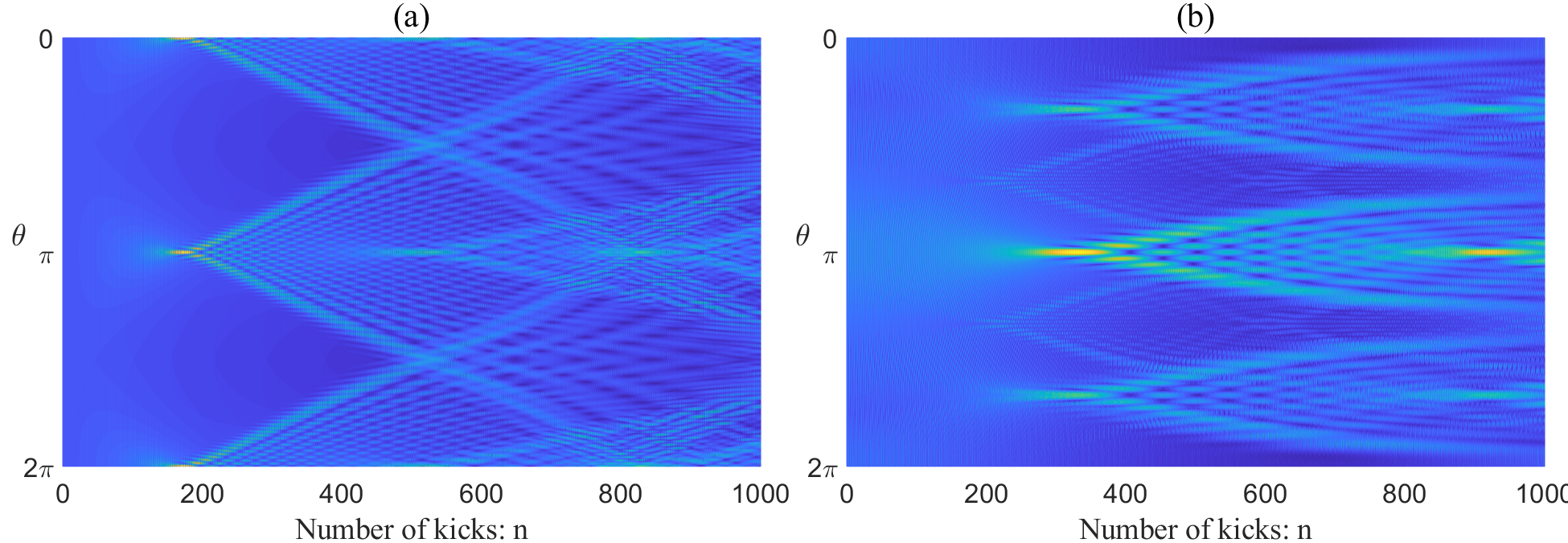}
\caption{Evolution of the wave amplitude $|\psi(\theta_{n},t_{n})|$ for the QKR near high-order resonance with detuning parameter $\Delta=0.0001$, initialized in a plane-wave state $\psi(\theta_{0},t_{0})=1/\sqrt{2\pi}$. Parameters: (a) $K=100,~T=2\pi+\Delta$; (b) $K=1,~T=4\pi/3+\Delta$.}
\label{Fig-Quantum Caustics near HOR}
\end{figure}

For a plane wave with non-zero initial momentum, the quantum evolution is shown in Fig.~\ref{Fig-Quantum Caustics with Non-zero Initial Momentum}(a). The results demonstrate that although caustics still form and retain their characteristic features, the cusp positions exhibit periodic oscillations. Additionally, we simulate a Gaussian wave packet initially centered at $\pi/2$ with a standard deviation of 1, as shown in Fig.~\ref{Fig-Quantum Caustics with Non-zero Initial Momentum}(b). Unlike the plane-wave case, this spatially localized wave packet exhibits negligible probability density in the $\pi$ to $2\pi$ range, suppressing the cusp caustic formation while preserving both the fundamental interference structure and some of its fold structure.

Furthermore, numerical simulations reveal that similar caustic phenomena also emerge in the near-high-order-resonant regimes, where the kicking period satisfies $T=4\pi r/s+\Delta$ for coprime integers $r$ and $s$ with $s \ge 2$, as shown in Fig.~\ref{Fig-Quantum Caustics near HOR}. Notably, in addition to the existence of cusp caustics in the vicinity of $\theta=\pi$, secondary caustic structures characterized by weaker intensity yet distinct features arise at other spatial positions, with the formation of such secondary caustic structures displaying a marked dependence on the kicking strength. In addition, we check the robustness of the above caustic structures subject to multiplicative Gaussian white noise~\cite{Wiseman2009}, and confirm that the caustic structures persist under low-intensity noise.

In the following section, taking the near-primary-resonant case for instance, we present a theoretical analysis of the mechanisms underlying these caustics.

\section{Theoretical Analysis}\label{sec-3}
\subsection{Path Integral Method}
Since our model is time-periodic, Floquet theory implies that the evolution operator over one period can be expressed as $\hat{U}(T)=\hat{\mathcal{T}}\exp\left(-i\int_{0}^{T}\hat{H}dt\right)=\exp(-iK\cos\hat{\theta})\exp\left(-i\hat{p}^2T/2\right)$, where $\hat{\mathcal{T}}$ is the time-ordering operator. For the discrete-time series $\{t_{j}\}$ with uniform spacing $t_{j+1}-t_{j}=T$, the evolution operator reduces to
\begin{equation}\label{Evolution Operator}
\hat{U}(t_{j+1};t_{j})=\exp\left(-iK\cos\hat{\theta}\right)\exp\left(-i\frac{\hat{p}^2}{2}\Delta\right).
\end{equation}
To characterize the quantum dynamical behavior, we employ the Feynman path integral formalism \cite{Feynman2010}. The dynamical evolution of the initial quantum state $\psi(\theta_0,t_0)$ to time $t_n$ is described by
\begin{equation}\label{wave function}
\psi(\theta_{n},t_{n})=\int_{0}^{2\pi} G(\theta_{n},t_{n};\theta_0,t_{0})\psi(\theta_{0},t_0)d\theta_{0},
\end{equation}
where the propagator $G(\theta_{n},t_{n};\theta_0,t_{0})$ takes the form,
\begin{equation}
G(\theta_{n},t_{n};\theta_0,t_{0})=\bigg\langle \theta_{n}\bigg|\left(\prod_{j=0}^{n-1}\hat{U}(t_{j+1};t_{j})\right)\bigg|\theta_{0}\bigg\rangle.
\end{equation}
For deriving the path integral representation, we express the propagator in the angle basis through insertions of the identity operator $\int d\theta_{k} \ket{\theta_{k}}\bra{\theta_{k}}=1$, where $d\theta_{k}$ denotes the angular position volume element at each intermediate time $t_{k}$ (from $k=1$ to $n-1$). The resulting propagator is then given by 
\begin{equation}
G(\theta_{n},t_{n};\theta_0,t_{0})=\int\prod_{k=1}^{n-1}d\theta_{k}\cdot\prod_{j=0}^{n-1}\bra{\theta_{j+1}}\hat{U}(t_{j+1};t_{j})\ket{\theta_{j}}.
\end{equation}
The matrix element $\bra{\theta_{j+1}}\hat{U}(t_{j+1};t_{j})\ket{\theta_{j}}\equiv M_{j+1,j}$ can be represented by inserting a complete set of angular momentum eigenstates,
\begin{equation}
\begin{split}
&M_{j+1,j}=\sum_{p_{j}=-\infty}^{\infty}\langle\theta_{j+1}|\hat{U}(t_{j+1};t_{j})\ket{p_{j}}\bra{p_{j}} \theta_{j}\rangle\\
=&\frac{1}{2\pi}e^{-iK\cos\theta_{j+1}}\sum_{p_{j}=-\infty}^{\infty}\exp\left[-i\frac{p_{j}^2}{2}\Delta+i(\theta_{j+1}-\theta_{j})p_{j}\right]\\
\equiv&\frac{1}{2\pi}\exp\left(-iK\cos\theta_{j+1}\right)\vartheta_{3}\left(\frac{\theta_{j+1}-\theta_{j}}{2}, \exp\left(-i\frac{\Delta}{2}\right)\right).
\end{split}
\end{equation}
Here, the standard angular coordinate-momentum representation $\langle \theta_{j} | p_{j}\rangle=\frac{1}{\sqrt{2\pi}}e^{i\theta_{j}p_{j}}$ has been used, and $\vartheta_{3}(z, q)$ denotes the Jacobi theta function~\cite{Whittaker2021}. To proceed, we apply the Poisson summation formula to transform the Jacobi theta function as: $\vartheta_{3}\left(\frac{\theta_{j+1}-\theta_{j}}{2},e^{-i\frac{\Delta}{2}}\right)=\sqrt{\frac{2\pi}{i\Delta}}\sum_{k=-\infty}^{\infty}\exp\left[i\frac{(\theta_{j+1}-\theta_{j}-2\pi k)^2}{2\Delta}\right]$. In the $\Delta\to0$ limit, where only the $k=0$ term dominates, the matrix element consequently reduces to
\begin{equation}
M_{j+1,j}=\sqrt{\frac{1}{2\pi i\Delta}}\exp\left[i\frac{(\theta_{j+1}-\theta_{j})^2}{2\Delta}-iK\cos\theta_{j+1}\right].
\end{equation}
Now, the propagator can be written as
\begin{equation}\label{Propagator}
G(\theta_{n},t_{n};\theta_0,t_{0})=\int\mathcal{D}\theta\exp\bigg\{i\sum_{j=0}^{n-1}\bigg[\frac{(\theta_{j+1}-\theta_{j})^2}{2\Delta}-K\cos\theta_{j+1}\bigg]\bigg\}\equiv\int\mathcal{D}\theta\exp\left(iS\right),
\end{equation}
where $\mathcal{D}\theta$ denotes the path integral measure defined by $\sqrt{\frac{1}{2\pi i\Delta}}\prod_{k=1}^{n-1}\sqrt{\frac{1}{2\pi i\Delta}}d\theta_{k}$, and $S$ represents the discrete action.

\subsection{Semiclassical Caustic Trajectory}
\begin{figure}[t]
\centering
\includegraphics[width=1\linewidth]{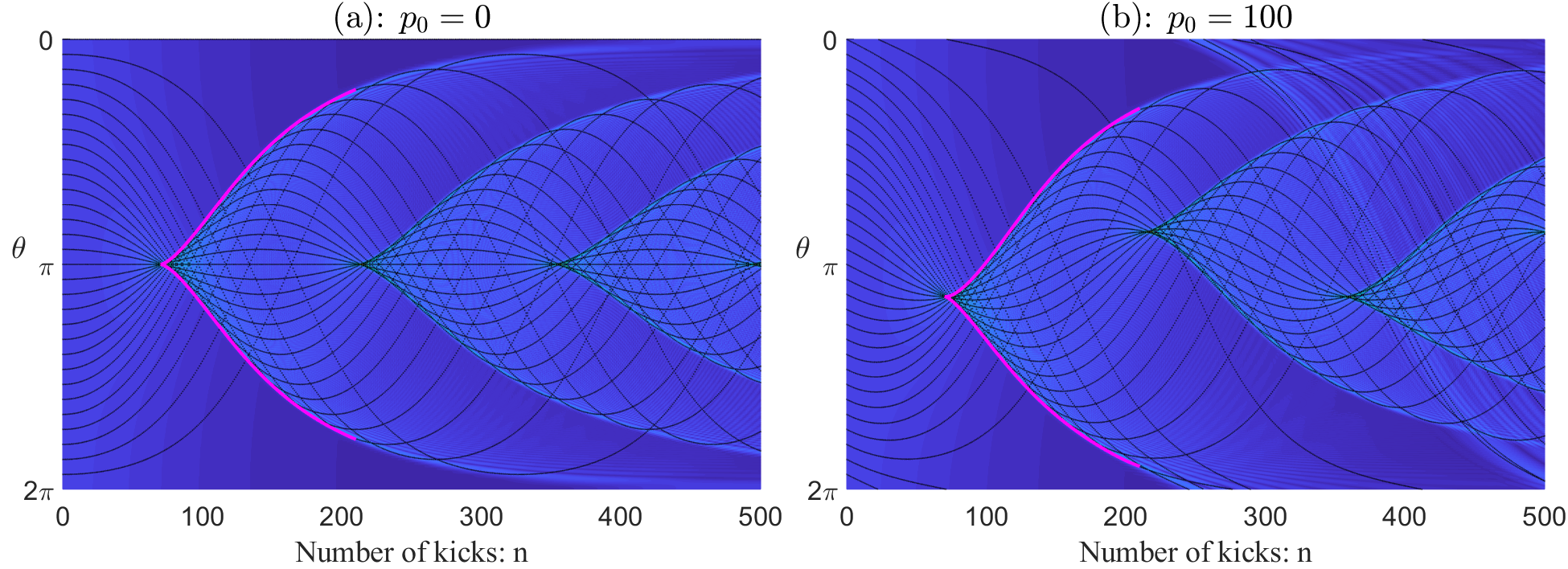}
\caption{Dynamical evolution of the QKR and its corresponding classical mapping in Eq.~(\ref{Equivalent Mapping}). Surface plots depict the evolution of $|\psi(\theta_n,t_n)|$ with the initial plane-wave state $\psi(\theta_0,t_0)=(1/\sqrt{2\pi})e^{ip_{0}\theta_{0}}$. The black solid lines represent numerical simulations of the classical mapping with initial momentum $p_{0}^{\text{cl}}=p_{0}\Delta$. The magenta solid lines correspond to the caustic curve of the first caustic. Other parameters are $K=5$ and $\Delta=0.0001$.}
\label{Fig-Classical Paths}
\end{figure}

Within the Feynman path integral framework, for small detuning $\Delta$, the dominant contribution to the propagator arises from the classical path $\{\theta_{j}^{\text{cl}}\}$, and quantum effects emerge as perturbative corrections around this extremal path~\cite{Feynman2010, Berry1972}. Employing a semiclassical expansion about the classical path, we decompose the path as $\theta_{j}=\theta_{j}^{\text{cl}}+\eta_{j}$, where $\eta_{j}$ represents small quantum fluctuations that satisfy the Dirichlet boundary conditions $\eta_{0}=\eta_{n}=0$. The discrete action $S$ can be expanded in a Taylor series about the classical path:
\begin{equation}
S[\theta]=S[\theta^{\text{cl}}]+\frac{1}{2}\sum_{j,k=1}^{n-1}\frac{\partial^2 S}{\partial \theta_{j}\partial \theta_{k}}\bigg|_{\theta_{j}^{\text{cl}},\theta_{k}^{\text{cl}}}\eta_{j}\eta_{k}+\mathcal{O}(\eta^3),
\end{equation}
where the quadratic term characterizes Gaussian fluctuations around the classical path, and higher-order terms account for non-Gaussian quantum corrections. In this subsection, we restrict our analysis to classical paths, with quantum corrections to be systematically addressed later. This semiclassical expansion approach requires the action to be extremized with respect to all $\theta_{j}$, leading to the discrete-time Euler-Lagrange equation:
\begin{equation}\label{Euler-Lagrange Equation}
\frac{\theta_{j+1}^{\text{cl}}-2\theta_{j}^{\text{cl}}+\theta_{j-1}^{\text{cl}}}{\Delta}=K\sin\theta_{j}^{\text{cl}},~~j=1,2,3...
\end{equation}
which corresponds to the discrete mapping 
\begin{subequations}\label{Equivalent Mapping}
\begin{align}
\theta_{j+1}^{\text{cl}}=&\theta_{j}^{\text{cl}}+p_{j}^{\text{cl}}~~(\text{mod}~2\pi),\\
p_{j+1}^{\text{cl}}=&p_{j}^{\text{cl}}+K\Delta\sin\theta_{j+1}^{\text{cl}},
\end{align}
\end{subequations}
where $p_{0}^{\text{cl}}\equiv p_{0}\Delta$ denotes the effective classical momentum, scaled by the detuning parameter, and $\theta_{0}^{\text{cl}}\equiv\theta_{0}$ is the corresponding classical angular variable. The similar classical mapping in the near-resonant regime has also been derived in Refs.~\cite{Wimberger2004, Wang2011, Zou2024}, with the help of a fictitious classical limit. Our deduction here is based on the classical-quantum correspondence of the path integral method. As shown in Fig.~\ref{Fig-Classical Paths}, classical paths with initial angles uniformly distributed over $[0,2\pi)$ accurately reproduce the quantum caustic structure.

\begin{figure}[t]
\centering
\includegraphics[width=1\linewidth]{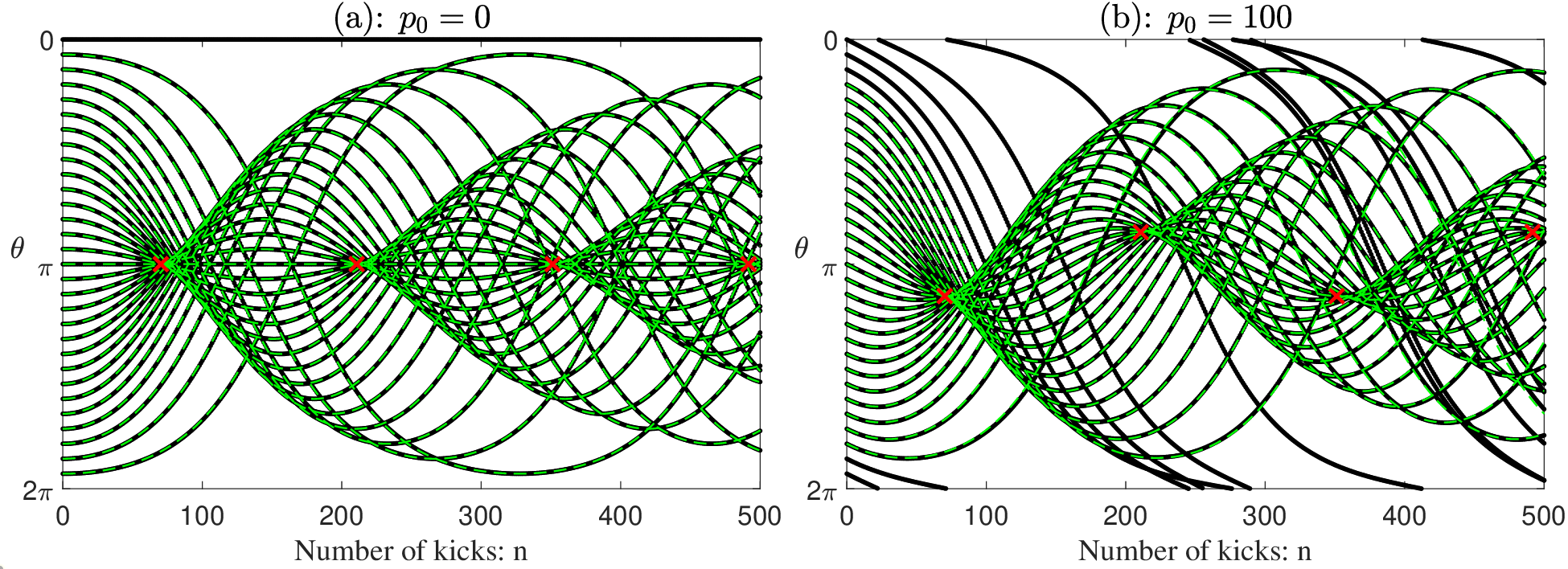}
\caption{Comparison between classical discrete mapping and continuum analytical solutions. The black solid and green dashed lines correspond to Eq.~(\ref{Equivalent Mapping}) and Eq.~(\ref{Analytical Solution}), respectively. The red crosses mark the cusp points from Eqs.~(\ref{Caustic Time}) and (\ref{Cusp Points}). Parameters are identical to those in Fig.~\ref{Fig-Classical Paths}.}
\label{Fig-Analytical Solutions}
\end{figure}

In the continuum limit $\Delta\to0$ (i.e., the near-resonant condition), the discrete sequence $\{\theta_{j}^{\text{cl}}\}$ converges to a continuous trajectory $\theta_{\text{con}}^{\text{cl}}(t)$ with $t=j\Delta$. Under this approximation, the Euler-Lagrange equation (\ref{Euler-Lagrange Equation}) then reduces to the nonlinear pendulum equation:
\begin{equation}\label{Semiclassical Equation}
\ddot{\theta}_{\text{con}}^{\text{cl}}=\frac{K}{\Delta}\sin\theta_{\text{con}}^{\text{cl}}.
\end{equation}
Here, the overdot denotes differentiation with respect to time $t$ and initial momentum is $\dot{\theta}_{\text{con}}^{\text{cl}}(t=0)=p_0$. When the initial condition satisfies $\cos(\theta_0^{\text{cl}}-\pi)>p_0^2\Delta/2K-1$ (i.e., total energy $E<K/\Delta$), the system enters a bounded oscillatory state. In this regime, the phase-space trajectories exhibit characteristic closed elliptical features, corresponding to periodic libration dynamics. Conversely, if this condition is not satisfied, the system transitions to an unbounded rotational state, characterized by open phase-space trajectories. Such motion does not generate cusp caustic structures. Therefore, we only focus on the libration solution
\begin{equation}\label{Analytical Solution}
\theta_{\text{con}}^{\text{cl}}(t)=\pi+2\arcsin\left[k\cdot\text{sn}\left(t\sqrt{\frac{K}{\Delta}}+F(\varphi,k),k\right)\right],
\end{equation}
where $k=\sqrt{p_{0}^{2}\Delta/4K+m^2},~\varphi=\arcsin(m/k)$ and $m=\sin[(\theta_{0}^{\text{cl}}-\pi)/2]$. Here, $F(\varphi,k)$ represents the incomplete elliptic integral of the first kind and $\text{sn}(u,k)$ denotes the Jacobi elliptic sine function. Figure~\ref{Fig-Analytical Solutions} presents a comparison between the discrete mapping simulations and the continuum analytical solution given by Eq.~(\ref{Analytical Solution}) at sampling times $t=n\Delta$. As clearly demonstrated, excellent agreement is observed between the two solutions.

In the classical dynamical system, caustics emerge as fundamental singularities in phase space where families of classical trajectories focus. Mathematically, these singular points are characterized by the vanishing derivative condition ${\partial \theta_{\text{con}}^{\text{cl}}}/{\partial \theta_0^{\text{cl}}}=0$~\cite{Averbukh2001}. Although the exact analytical solution to this equation remains difficult to derive, it can be clearly observed from Fig.~\ref{Fig-Classical Paths} that the cusp caustics are primarily contributed by trajectories with initial angles distributed around $\pi$. To quantify this local structure, we perform a small-angle expansion of $\theta_{0}^{\text{cl}}$ around $\pi$ by setting $\theta_{0}^{\text{cl}}=\pi+\delta_{0}$, and retain terms up to the first order of $\delta_{0}$ in Eq.~(\ref{Analytical Solution}), yielding:
\begin{equation}\label{Approximate Solution}
\theta_{\text{con}}^{\text{cl}}(t)\approx\pi+\delta_{0}\cos\left(t\sqrt{\frac{K}{\Delta}}\right)+p_{0}\sqrt{\frac{\Delta}{K}}\sin\left(t\sqrt{\frac{K}{\Delta}}\right).
\end{equation}
Differentiating this approximate solution with respect to $\delta_{0}$ and setting the derivative to zero yields the caustic time (i.e., the time of cusp points):
\begin{equation}\label{Caustic Time}
n_{\text{cau}}=\frac{t}{\Delta}=\frac{(2m+1)\pi}{2\sqrt{K\Delta}},~~m=0,1,2,\dots.
\end{equation}
Notably, within the first-order approximation, this solution is independent of $p_{0}$, implying that the occurrence time of cusp points is unaffected by the initial momentum under this level of approximation. This characteristic provides a simplified yet insightful perspective on the temporal behavior of the cusp caustic. Substituting this solution back into Eq.~(\ref{Approximate Solution}) yields the angles where the cusp points occur:
\begin{equation}\label{Cusp Points}
\theta_{\text{cusp}}=\pi+(-1)^{m}p_{0}\sqrt{\frac{\Delta}{K}},~~m=0,1,2,\dots.
\end{equation}
This result is shown in Fig.~\ref{Fig-Analytical Solutions}, marked by red crosses. Moreover, we numerically solve the caustic equation ${\partial \theta_{\text{con}}^{\text{cl}}}/{\partial \theta_0^{\text{cl}}}=0$ using the analytical solution in Eq.~(\ref{Analytical Solution}), then evaluate the system dynamics by substituting these critical points back into the same governing equation for final characterization. The resultant profiles, displayed as magenta solid curves in Fig.~\ref{Fig-Classical Paths}, not only capture the characteristic features of classical caustics but also exactly reproduce the quantum caustic structures. This operational procedure confirms that the quantum caustic dynamics can be completely reconstructed through these classical solutions.

\subsection{Quantum Caustics and Scaling Law}
The semiclassical analysis establishes a clear correspondence between theoretically derived semiclassical caustics and their numerically observed quantum manifestations. We now demonstrate how these quantum caustics emerge fundamentally from the underlying propagator structure. In the continuum limit $\Delta\to0$ under stationary phase approximation, the quantum propagator takes the following form
\begin{equation}
G(\theta,t;\theta_{0},t_{0})=e^{iS[\theta_{\text{con}}^{\text{cl}}]}\int\mathcal{D}\eta e^{\frac{i}{2}\delta^2 S[\theta_{\text{con}}^{\text{cl}}]},
\end{equation}
where 
\begin{equation}
\begin{split}
\delta^2S[\theta_{\text{con}}^{\text{cl}}]=&\int_{t_{0}}^{t}dt^\prime\left(\dot{\eta}^2+\frac{K}{\Delta}\cos\theta_{\text{con}}^{\text{cl}}\eta^2\right),
\end{split}
\end{equation}
and a continuum functional measure
\begin{equation}
\mathcal{D}\eta=\lim_{\Delta\to 0}\sqrt{\frac{1}{2\pi i\Delta}}\prod_{t^\prime=t_0}^{t}\sqrt{\frac{1}{2\pi i\Delta}}d\eta(t^\prime).
\end{equation}
Applying integration by parts to the action variation, we can express $\delta^2 S[\theta_{\text{con}}^{\text{cl}}]$ in a compact quadratic form
\begin{equation}
\delta^2S[\theta_{\text{con}}^{\text{cl}}]=\int_{t_{0}}^{t}dt^\prime\eta\hat{O}\eta,
\end{equation}
where the operator $\hat{O}$ is defined as
\begin{equation}
\hat{O}=-\frac{d^2}{dt^{\prime2}}+\frac{K}{\Delta}\cos\theta_{\text{con}}^{\text{cl}}.
\end{equation}
Finally, using the Gaussian functional integral formula, the propagator simplifies to
\begin{equation}\label{C-Propagator}
G(\theta,t;\theta_{0},t_{0})\propto e^{iS[\theta_{\text{con}}^{\text{cl}}]}(\det\hat{O})^{-1/2}.
\end{equation}
The functional determinant $\det\hat{O}$ for this second-order differential operator can be computed using the Gelfand-Yaglom formula~\cite{Gelfand1960, Dowker2012, Ossipov2018}, which requires solving the following initial value problem:
\begin{equation}
\hat{O}u(t)=0, \quad u(t_0)=0,~~\dot{u}(t_0)=1.   
\end{equation}
The determinant is then given by
\begin{equation}
\det\hat{O}=u(t).  
\end{equation}
By differentiating the classical equation of motion Eq.~(\ref{Semiclassical Equation}) with respect to the initial condition $\theta_0^{\text{cl}}$, we obtain the quantity $u(t)={\partial \theta_{\text{con}}^{\text{cl}}}/{\partial \theta_0^{\text{cl}}}$, which becomes singular at caustics and demonstrates the classical-quantum correspondence. Importantly, the condition $\det\hat{O}\to0$ not only signals the failure of the stationary phase approximation but also reflects the emergence of structural singularities in the propagator's phase space representation. These singularities amplify quantum fluctuations, leading to strongly localized probability densities with delta-function-like peaks, which are a signature of quantum caustic structures.
\begin{figure}[t]
\centering
\includegraphics[width=\linewidth]{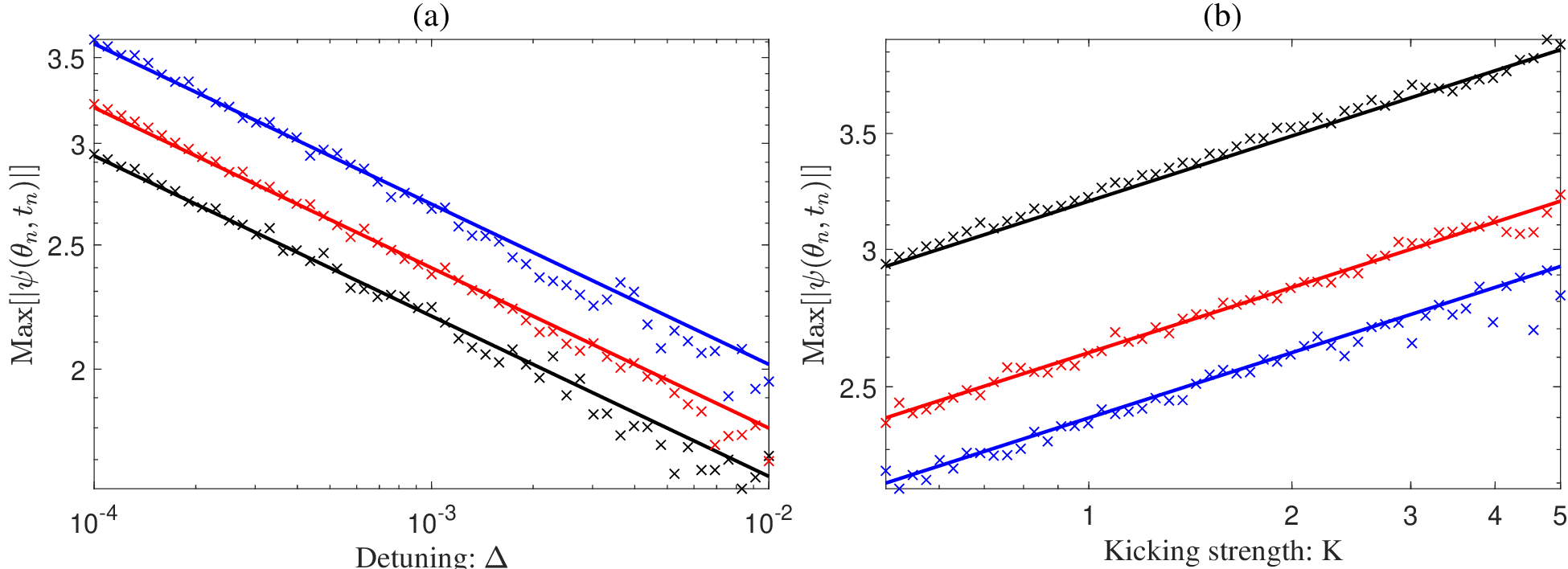}
\caption{Scaling behavior of wave amplitude $|\psi(\theta_{n},t_{n})|$ at the first cusp point. The solid lines represent $2|\lambda|^{1/4}$ where $\lambda=-\frac{\pi}{48}\sqrt{\frac{K}{\Delta}}$ and $2$ is a fitting value. The crosses mark the numerical simulation results. The system is initialized with a plane wave $\psi(\theta_{0},t_{0})=(1/\sqrt{2\pi})e^{ip_{0}\theta_{0}}$, where $p_{0}=(\pi/4)\sqrt{K/\Delta}$. (a) From top to bottom, $K=2.5,~1,~0.5$; (b) From top to bottom, $\Delta=0.0001,~0.0005,~0.001$.}
\label{Fig-Scaling Law}
\end{figure}

The wave function in the continuum limit is obtained by substituting Eq.~(\ref{C-Propagator}) into Eq.~(\ref{wave function}) with the initial condition $\psi(\theta_{0},t_{0})=1/\sqrt{2\pi}e^{ip_{0}\theta_{0}}$:
\begin{equation}\label{C-wave function}
\psi(\theta,t)=\mathcal{C}\int_{0}^{2\pi}e^{iS[\theta_{\text{con}}^{\text{cl}}]}e^{ip_{0}\theta_{0}}d\theta_{0},
\end{equation}
where $\mathcal{C}$ is a time-dependent normalization factor and the continuum action is 
\begin{equation}
S[\theta_{\text{con}}^{\text{cl}}]=\int_{t_{0}}^{t}dt^\prime\left[\frac{(\dot{\theta}_{\text{con}}^{\text{cl}})^2}{2}-\frac{K}{\Delta}\cos\theta_{\text{con}}^{\text{cl}}+K\dot{\theta}_{\text{con}}^{\text{cl}}\sin\theta_{\text{con}}^{\text{cl}}\right].
\end{equation}
Previous studies have shown that when the initial momentum $p_{0}\ne 0$, the cusp points oscillate periodically around $\pi$. To characterize the dynamical behavior of the system near $\pi$, we introduce the transformations $\theta_{\text{con}}^{\text{cl}}=\pi+\delta$ and $\theta_{0}^{\text{cl}}=\pi+\delta_{0}$. By substituting these transformations into the equation of motion given in Eq.~(\ref{Semiclassical Equation}), the wave function in Eq.~(\ref{C-wave function}) can be expressed in the following form:
\begin{equation}\label{Eq26}
\begin{split}
&\psi(\theta,t)=\mathcal{C}\int_{-\pi}^{\pi}d\delta_{0}\exp\bigg\{i\bigg[\frac{p_{0}^{2}}{2}+\frac{2K}{\Delta}\cos\delta-\frac{K}{\Delta}\cos\delta_{0}\\
+&K\sin\delta\sqrt{p_{0}^{2}+\frac{2K}{\Delta}(\cos\delta-\cos\delta_{0})}\bigg](t-t_{0})+ip_{0}(\pi+\delta_{0})\bigg\}.
\end{split}
\end{equation}
Here, $\delta$ and $t$ are regarded as control parameters rather than variables. Expanding $\cos\delta_0$ in a Taylor series around $\delta_0=0$ up to $\mathcal{O}(\delta_0^4)$ yields $\cos\delta_0=1-\frac{1}{2}\delta_0^2+\frac{1}{24}\delta_0^4$. From Eq.~(\ref{Eq26}), the coefficient of $\delta_0^4$ is determined as $-i\frac{K}{\Delta}\frac{t-t_{0}}{24}\equiv i\lambda$, where a small correction term originating from the higher-order expansion of the square root term has been neglected. Considering the behavior near the first cusp point, we set $t-t_{0}=\frac{\pi}{2}\sqrt{\frac{\Delta}{K}}$, so that $\lambda=-\frac{\pi}{48}\sqrt{\frac{K}{\Delta}}$. Based on the framework of catastrophe theory, near the cusp point, the wave amplitude follows a power-law scaling behavior~\cite{Berry1977, Arnold1992, Kirkby2019, Mumford2019}:
\begin{equation}
|\psi(\theta,t)|\propto|\lambda|^{1/4}\propto\left(\frac{K}{\Delta}\right)^{1/8},
\end{equation}
where the exponent $1/4$ is Arnold index of cusp catastrophe. This behavior is valid solely when the oscillation amplitude of the cusp is sufficiently small. Specifically, this condition is satisfied when $\theta_{\text{cusp}}$ in Eq.~(\ref{Cusp Points}) lies close to $\pi$, such that the libration solution dominates. We take the maximum amplitude of the wave function to approximate the value at the cusp point for numerical simulations where the cusp oscillation amplitude is limited to $\pi/4$. The comparison between these results and the analytical scaling law is shown in Fig.~\ref{Fig-Scaling Law}.

\subsection{Chaos Disrupts Caustics}
\begin{figure*}[!ht]
\centering
\includegraphics[width=1\linewidth]{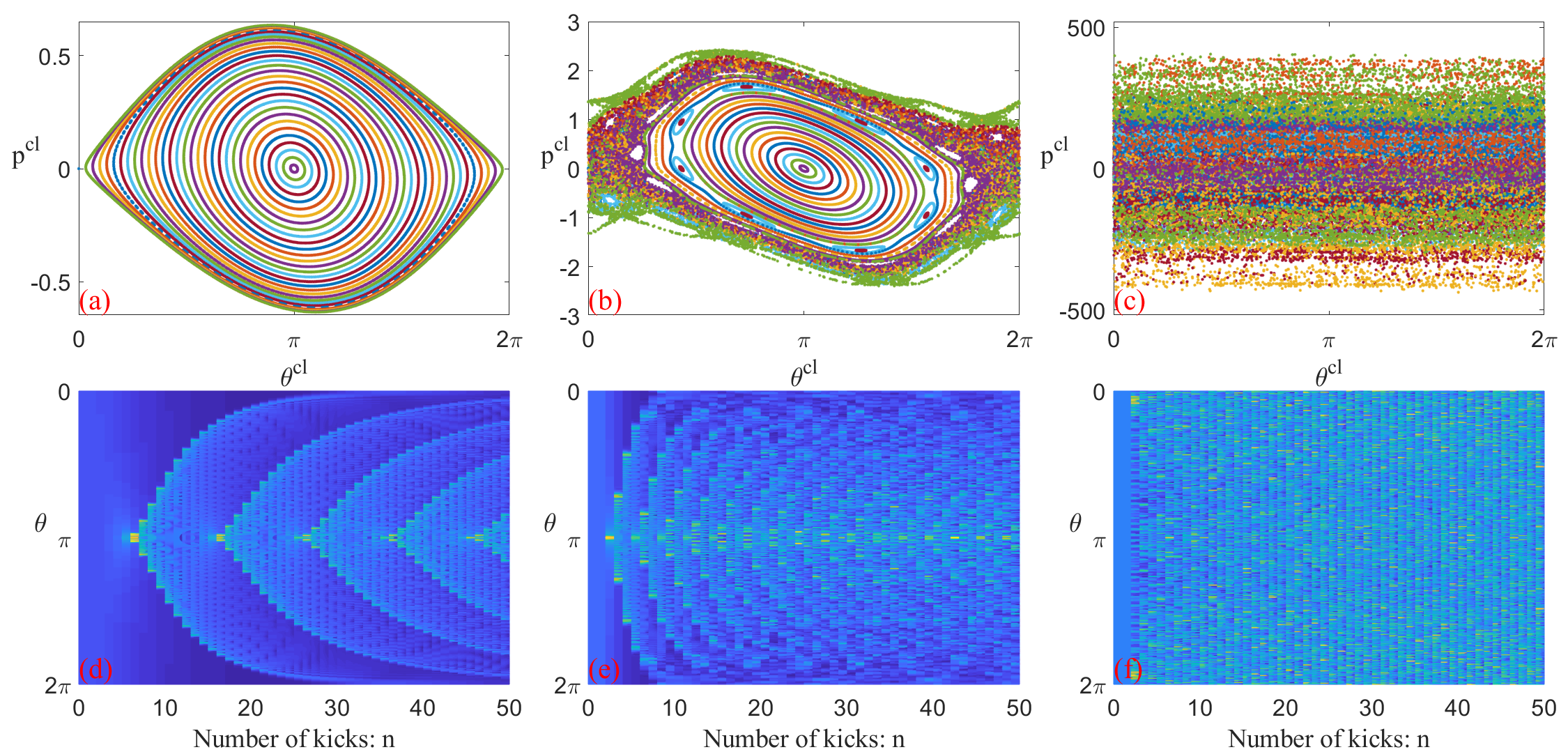}
\caption{Phase diagram of the discrete mapping in Eq.~(\ref{Equivalent Mapping}) (top) and corresponding quantum evolution of the QKR (bottom) at a fixed kicking strength $K=100$. The initial states for both classical and quantum cases are identical to those in Fig.~\ref{Fig-Classical Paths}(a). (a, d) $\Delta=0.1/K$; (b, e) $\Delta=0.9716/K$; (c, f) $\Delta=5/K$.}
\label{Fig-Phase Diagram}
\end{figure*}
In this subsection, we take the special case of $p_{0}=0$ to qualitatively discuss how chaos disrupts caustic structures. Applying the variable transformation $\theta_{\text{con}}^{\text{cl}}=\pi+\delta$, the equation of motion Eq.~(\ref{Semiclassical Equation}) becomes $\ddot{\delta}+(K/\Delta)\sin\delta=0$. In the small-angle limit $|\delta|\ll1$, the sine function linearizes to $\sin\delta\approx\delta$. This linearization yields a simple harmonic oscillator, $\ddot{\delta}+(K/\Delta)\delta=0$, with an intrinsic oscillation period $T=2\pi\sqrt{\Delta/K}$ that is independent of initial conditions. For initial states localized within a neighborhood of $\theta_0^{\text{cl}}=\pi$, the linearized solution predicts periodic focusing at discrete-time instances $t=(2m+1)\pi/(2\sqrt{K/\Delta})$ (with $t=n\Delta$). This simple picture establishes a correlation between caustic structures and Kolmogorov-Arnold-Moser invariant tori in phase space as shown in Figs.~\ref{Fig-Phase Diagram}(a) and (d).
When the last Kolmogorov-Arnold-Moser invariant torus breaks at $K\Delta > 0.9716$~\cite{Chirikov1979}, global chaos emerges as illustrated in Fig.~\ref{Fig-Phase Diagram}(b). We observe the attenuation of quantum caustic structures in Fig.~\ref{Fig-Phase Diagram}(e). A further increase in 
$K\Delta$ to 5 drives the system into a state of global fully developed chaos~\cite{Santhanam2022}, marked by the complete disappearance of regular phase-space structures and the total vanishing of quantum caustic structures as shown in Fig.~\ref{Fig-Phase Diagram}(f). 

\section{Conclusion}\label{sec-4}
In conclusion, we have thoroughly investigated the dynamics of the celebrated QKR model and revealed the striking recurring cusp caustics in the near-resonant regime. Within a path integral framework and with the aid of the Gelfand-Yaglom method, we have obtained the analytical expression for the recurring cusp period and the position of the caustic patterns. Furthermore, we have established a scaling law of the amplitude at the first cusp point through catastrophe theory. The influence of chaos on these phenomena has also been discussed. 

The QKR model is a fundamental model that has been experimentally realized in various systems, such as optical systems~\cite{Rosen2000, Fischer2000, Zhang2015}, ultracold atoms~\cite{Moore1995, Chabe2008, Lemarie2009} and femtosecond laser-controlled molecular polarization~\cite{Johannes2015, Bitter2016}. Our findings could have implications for quantum control and simulation, and are expected to stimulate further experimental studies.

\backmatter
\bmhead{Acknowledgements}
The authors thank Wenlei Zhao and Yu Chen for helpful discussions. This work was supported by the National Natural Science Foundation of China (Grant No.~U2330401).

\bibliography{refs}

\end{document}